# Band-selective Measurement of Electronic Dynamics in $VO_2$ using Femtosecond Near Edge X-ray Absorption.

A. Cavalleri[(1)*], M. Rini[(1)], H.H.W. Chong[(1)], S. Fourmaux[(3)], T.E. Glover[(2)], P.A. Heimann[(2)], J.C. Kieffer[(3)], R.W. Schoenlein[(1)].

[(1)]*Materials Sciences Division Lawrence Berkeley National Laboratory.*
[(2)]*Advanced Light Source Division, Lawrence Berkeley National Laboratory.*
[(3)]*Université du Québec, INRS énergie et matériaux, Varennes, Québec.*

We report on the first demonstration of femtosecond x-ray absorption spectroscopy and on the first scientific application of femtosecond synchrotron radiation. Our experiments are made possible by the use of broadly tunable bending-magnet radiation from "laser sliced" electron bunches within a synchrotron storage ring. We measure ultrafast electronic dynamics associated with the photo-induced insulator-metal phase transition in $VO_2$. Symmetry- and element–specific x-ray absorption from near-500-eV V2p and O1s core levels separately measures the filling dynamics of differently hybridized V3d-O2p electronic bands near the Fermi level.

X-ray measurements on the femtosecond timescale combine powerful probes of matter with access to elementary timescales of atomic motion. Scientific applications in this area have to date concentrated almost exclusively on time-resolved diffraction with hard x-ray pulses[1,2,3,4,5]. These experiments have been driven in large part by developments in monochromatic tabletop plasma sources[6], x-ray generation by femtosecond-laser interaction with relativistic electron-beams[7], and linac-based ultra-fast x-ray pulses[8]. Recent breakthroughs have also opened the way to studies of femtosecond atomic-structural dynamics with electron diffraction[9].

X-ray absorption techniques are important complements to diffraction, probing local electronic, magnetic and short-range atomic structures with element specificity. A long-standing scientific frontier is their application in time-resolved experiments with femtosecond resolution, to interrogate the dynamics of electronic and magnetic phase transitions in solids or the atomic motions of chemical processes in disordered liquids. Yet, because most of the existing femtosecond x-ray sources are not easily tuned, this class of experiments has to date remained an elusive goal.

In this paper we report on the first demonstration of femtosecond x-ray absorption spectroscopy. Our experiments are made possible by the use of "laser-sliced" synchrotron pulses, which combine the broad tunability of bending magnet radiation with the femtosecond time duration of lasers[10]. We measure the electronic dynamics in $VO_2$, a compound that undergoes an insulator-to-metal phase transition when photo-excited with light[11]. Time-resolved near edge x-ray absorption spectroscopy at the symmetry-selective Vanadium L edge (516 eV) and at the Oxygen K edge (531 eV) allows separate access to the variously hybridized V3d and O2p orbitals near the Fermi level.

The photo-induced insulator-to-metal transition in $VO_2$ is one of a broader class of ultrafast photo-control phenomena in inorganic[12] and organic[13] correlated-electron systems that have recently raised interest due to their importance for both basic science and applications[14]. These phase transitions proceed along ultrafast, non-thermal pathways, whereby different degrees of freedom (structural, electronic, magnetic) interact dynamically along the non-equilibrium pathway of the phase transition. This scenario opens a new window on the physics of strong correlations in condensed matter, but it also requires direct, ultrafast measurements of both atomic and electronic structures. In its low-T phase $VO_2$ has a monoclinic structure and insulating characteristics, derived from the high-T metal through cell-doubling and charge localization. Photo-excitation of the insulator results in the depletion of these localized valence-band states, while electrons are injected into the spatially extended conduction band. As a result of prompt "photo-doping", (1) the structural dimerization of the low-T insulator is coherently relaxed and (2) the metallic phase is formed, both processes occurring on the 100-fs timescale[3,15].

Photo-induced formation of the metallic state of $VO_2$ is apparent from the time-resolved, infrared absorption measurements displayed in figure 1. In these pump-probe experiments, above-gap excitation with 800-nm pulses at 1-KHz was combined with transmission probing at various wavelengths in the near infrared. Above a threshold fluence of 3 $mJ/cm^2$, the infrared absorption coefficient exhibited a prompt, step-like increase toward that of an opaque metallic state. Above 10 $mJ/cm^2$, the changes in the optical properties were independent on excitation level and saturated for a broad fluence interval, extending to the damage threshold of 50 $mJ/cm^2$. Such saturated dependence on the excitation fluence, as well as the absence of relaxation on the picosecond timescale, rules out mere carrier excitation as an explanation for the observed dynamics. The data are rather indicative of an ultrafast insulator-to-metal transition, as concluded in previous experiments[3,14,15]. Formation of the metallic state is evidenced by (1) impulsive formation of a near-1-eV plasmon resonance and (2) collapse of the semiconducting bandgap at 0.67 eV. At longer time delays the metallic phase is stabilized by thermalization above $T_c$=340 K, followed by recovery of the insulating phase on the nanosecond timescale. Beyond the identification of a transient metallic phase, the electronic dynamics within this complex electronic structure cannot be exhaustively addressed with infrared spectroscopy alone. In fact, the response at visible and near-infrared wavelengths is only sensitive to the joint density of bandgap states and is moreover affected by the collective response of the electronic plasma.

The near-the-Fermi-level electronic states that participate in the ultrafast insulator-to-metal transition are the two lowest-lying t2g-like states of the V3d manifold, namely the $V3d_{//}$ and $V3d_{\pi}$ states (see figure 1). The $V3d_{//}$ states, derived from weakly hybridized Vanadium 3d orbitals, are split into valence and conduction band in the cell-doubled insulator, and merge across the Fermi level in the metal. Because of dipole selection rules, these 3d-symmetry bands are best probed by measuring x-ray absorption from the V2p core levels near 516 eV. The $V3d_{\pi}$ band originates instead from stronger mixing between V3d and O2p orbitals[16], forming one of the two conduction bands in the low-T insulator and becoming metallic band above $T_c$. Their important 2p character makes the $V3d_{\pi}$ states most visible at the 531-eV O1 resonance[17].

Time-resolved x-ray absorption measurements were performed with femtosecond pulses of synchrotron radiation, generated by laser modulation of the electron-bunch energy within a wiggler (see figure 2). The modulated time slice was spatially separated from the main orbit in the radiating bending magnet, where the electron beam energy was dispersed along the radius of the storage ring. Femtosecond radiation was thus emitted off axis by the bending magnet, with a characteristic spectrum that continuously extended between the THz to the hard x-ray region. The femtosecond pulses of synchrotron radiation, focused onto the $VO_2$ sample with a toroidally bent silicon mirror, were spatially selected using a slit in the image plane of the storage ring. The laser pulses used to impress energy modulation onto the electron bunches and those used to excite the sample were derived from the same oscillator, allowing for absolute synchronization between optical pump and x-ray probe. A flat-field imaging spectrometer was used to disperse the transmitted soft x-rays after the sample, generating femtosecond spectra between 100 eV and 800 eV, with a resolution of approximately 4 eV.

Static spectra are shown in figure 2, alongside a 100-meV-resolution measurement obtained with state-of-the art NEXAFS (beamline 6.3.2). Blue and red parts of the curve refer to transitions from the V2p and O1s into the unoccupied $V3d_{//}$ and $V3d_{\pi}$ states, respectively. A higher lying $3d_{\sigma}$ band is also visible as a separate peak only in the high-resolution O1s absorption resonance. This resonance corresponds to sigma-hybridized O2p-V3d orbitals from the third t2g-like state of the V3d manifold. The $3d_{\sigma}$ band does not take part in the phase transition and was not probed in our time resolved experiments, which concentrated instead on the rising edges of the $3d_{//}$ and $3d_{\pi}$ bands (see arrows in the spectrum of figure 2).

The measured femtosecond x-ray-absorption response is reported in figure 3. At the $V2p_{3/2}$ edge, a prompt increase in absorption was observed immediately after photo-excitation, recovering within a few picoseconds. At the O1s resonance, the absorption coefficient was also initially observed to increase, synchronously with that at the $V2p_{3/2}$ resonance. Enhanced O1s absorption at 530-eV was followed by bleaching and by relaxation on the same few-picosecond timescale as the signal at the 516-eV $V2p_{3/2}$ edge. Our interpretation of the data, schematically illustrated in figure 3, proceeds along the following lines. Two effects are important when measuring a photo-induced insulator-to-metal transition with femtosecond NEXAFS spectroscopy: (1) band-filling and bandgap collapse (2) dynamic shifting of the core levels.

First, photo-excitation depletes the d-symmetry valence band, while populating the $3d_\pi$ band. Photo-doping is immediately followed by bandgap collapse, and by the formation of a non-equilibrium metallic phase[18]. In this phase the electrons attain a transient, hot Fermi distribution with $T_e \gg T_l$, followed by electron-lattice thermalization within a few picoseconds. In the hot-electron metal, the broad $3d_\pi$ band is over-populated by the tail of the distribution, while the narrower $3d_{//}$ band has lower electron-occupancy than at equilibrium. For a rigid band structure and no shift in the core levels, a prompt increase in absorption at the V2p edge and bleaching at the O1s resonance is expected, with a decay time of few picoseconds. In the data, $V2p_{3/2}$ absorption promptly increases and indeed remains higher than at equilibrium for a few picoseconds, as expected from the considerations made above. After electron-lattice thermalization the transient increase in absorption recovers to the 1%-scale of the equilibrium phase change. This not discernible on the present measurements but it is clearly visible in higher-flux experiments with picosecond time resolution[19]. Thus, the 516-eV signal can be reconciled with the physical picture of figure 3. On the other hand, filling and shifting of the valence-bands alone do not explain the initial increase in absorption at the O1s edge, where mere bleaching would be expected.

Second, photo-excitation into the $3d_\pi$ band transiently increases the valency of the $O^{2-}$ anion, reducing that of the $V^{4+}$ cation. This effect could be as high as 0.25 electrons/$O^{2-}$ anion and 0.5 holes/ $V^{4+}$ cation, for the 50% photo-doping estimated at the 25 mJ/cm$^2$ used for the experiments. This valence-charge unbalance results in a dynamic shift of the O1s core level toward lower binding energy, while the V2p states shift toward higher binding energy. Thus, a blue-shift of the Vanadium L edge and a red-shift of the Oxygen K edge are expected immediately after excitation, driving x-ray absorption at 516-eV and at 530-eV in the opposite directions as non-equilibrium band filling. Competition between band filling and dynamic shifting of the core level binding energy is barely visible the V2p data, but it is clearly apparent at the O1s edge.

The shift of the V2p core levels is likely to be relatively small when compared to the 4-eV resolution of our experiments. This can be estimated by noting the small binding-energy differences for different oxides of Vanadium, amounting to few hundred meV for valency changes of one full electron/cation in $V_2O_3$ versus $VO_2$[20]. On the other hand, the effect on the O1s resonance is likely to be more pronounced. This can for instance be inferred from static binding energy measurements on $SiO_2$ and $ZrO_2$, where a 2-eV binding energy shift results from subtle differences in ionicity of the chemical bonding, despite a nominal $O^{2-}$ valency in both cases[21]. Significant improvement in our understanding of these processes will result from future experiments with improved spectrometer resolution and increased x-ray flux. Approximately one-thousand-fold increase in the femtosecond x-rays will result in the future by using undulator radiation and by increasing the laser repetition rate[22], making it possible to acquire entire x-ray absorption spectra with femtosecond time resolution.

In summary, we have demonstrated the first femtosecond x-ray-absorption spectroscopy experiment, made possible by the use of a femtosecond laser to manipulate the energy distribution of electron bunches within a synchrotron storage ring. The tunability of bending magnet radiation allows for femtosecond x-ray spectroscopy in the soft x-ray region below 1 keV, which is of great importance for the spectroscopy of transition metal oxides (L edges of metals and K edge of oxygen) and organics (K edge of Carbon), but difficult to access with any other femtosecond source. By comparing the response of photo-excited $VO_2$ at two absorption edges, we can separate the electronic dynamics within selected bands of different orbital symmetries. Ultrafast x-ray spectroscopy capabilities at any desired wavelength will, with appropriate flux increases, make possible new science in the areas of ultrafast magnetic phenomena, chemical dynamics at surfaces, in the gas phase or in liquids[23,24,25].

**Aknowledgements:** This work was supported by the Director, Office of Science, Office of Basic Energy Sciences, Division of Materials Sciences, of the U.S. Department of Energy under Contract No. DE-AC03-76SF00098. Work at INRS, Montreal was supported by NSERC and Canada Research Chair Program.

**Correspondence** should be addressed to Andrea Cavalleri: acavalleri@lbl.gov.

**FIGURE CAPTIONS**

**Figure 1: Electronic Structure of $VO_2$:** The near Fermi level bands that participate in the phase transition are the ligand-field-split, t2g-like V3d bands, namely the $3d_{//}$ and $3d_{\pi}$ bands. Above 340 K, the system becomes metallic, with the $3d_{//}$ band merging across the Fermi level and the $3d_{\pi}$ band shifting toward lower energies. **Femtosecond IR spectroscopy of the photo-induced insulator-metal transition.** The pump probe experiments were performed with 800-nm pulses for above-gap excitation at 25 mJ/cm$^2$ and IR probe pulses from an optical parametric amplifier. The samples used for the experiment were free-standing $VO_2$-$Si_3N_4$ (50-200 nm and 200-200 nm), obtained by sputtering of $VO_2$ on $Si_3N_4$–bulk-Si structure and chemical etching of the Silicon substrate.

**Figure 2: Static soft x-ray absorption of $VO_2$ at the V2p and O1s absorption resonances.** Soft x-ray absorption experiments were performed in transmission geometry on the same free-standing films used for the optical experiments. The high-resolution measurements (dashed line) are taken at beamline 6.3.2 of the *Advanced Light Source* with 100-meV resolution. Spectra with 4-eV resolution are measured at beamline 5.3.1, in the same geometry as the time-resolved experiments. **Femtosecond x-ray apparatus.** The sample was excited using 25 mJ/cm$^2$, 100-fs pulses form an amplified Ti:Sa laser operating at 800 nm, which was synchronized to the storage ring by a phase locked loop. The excitation pulse was obtained from the same oscillator used for slicing, resulting in absolute synchronization between sample-excitation pulses and femtosecond x-rays. The flux in the femtosecond x-rays was approximately four orders of magnitude lower than in the picosecond pulse, amounting to a few thousand photons/(sec 1% BW) at 500 eV in a 1-KHz train. Sample transmission (30%), spectrometer efficiency (6%) and detector efficiency (50%) resulted in about 10 photons/sec in the 1% bandwidth where the experiment was performed. The pulse duration of the laser-sliced synchrotron radiation was found to be less than 150 fs, as measured in the visible regime by optical cross-correlation with the excitation laser.

**Figure 3. Femtosecond XAS measurements at the V2p and O1s resonances. Data:** The experiments were performed by using the setup displayed in figure 2. The vertical error bars are determined by evaluating one standard deviation of the photon-counting distribution. The horizontal error bars are determined by evaluating the long-term drift in the delay between x-rays and excitation laser pulses. This drift (not corrected for in the present measurements) was evaluated over the several hour-long acquisition time by measuring the peak of the cross-correlation between the laser pulses and the visible-light emitted by the bending magnet. The continuous curves are guides to the eye. **Model.** Photo-doping rearranges electronic population between the charge-ordered, 3d-symmetry valence band and the mixed $3d_\pi$ orbitals. Bangap collapse results in the formation of a metal where the electrons are out of equilibrium with the lattice for 1-2 ps, maintaining charge unbalance until thermalization occurs. As a result of photo-excitation, the binding energy of the core levels is affected, with the V2p states becoming more bound and the O1s levels becoming less bound. The dynamics core-level shift recovers as the Fermi distribution reaches the lattice temperature and the valency of the equilibrium metal is re-established.

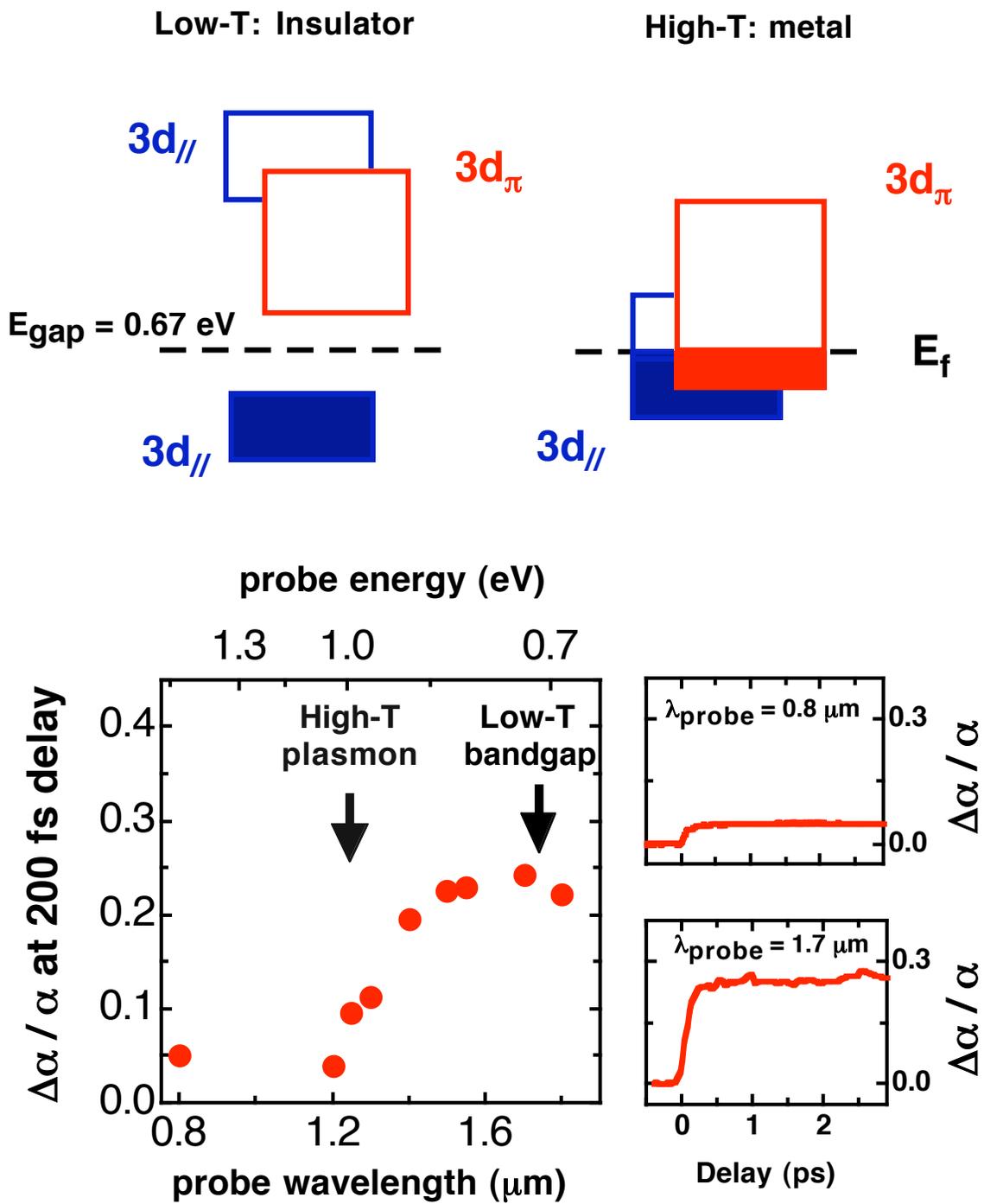

Figure 1

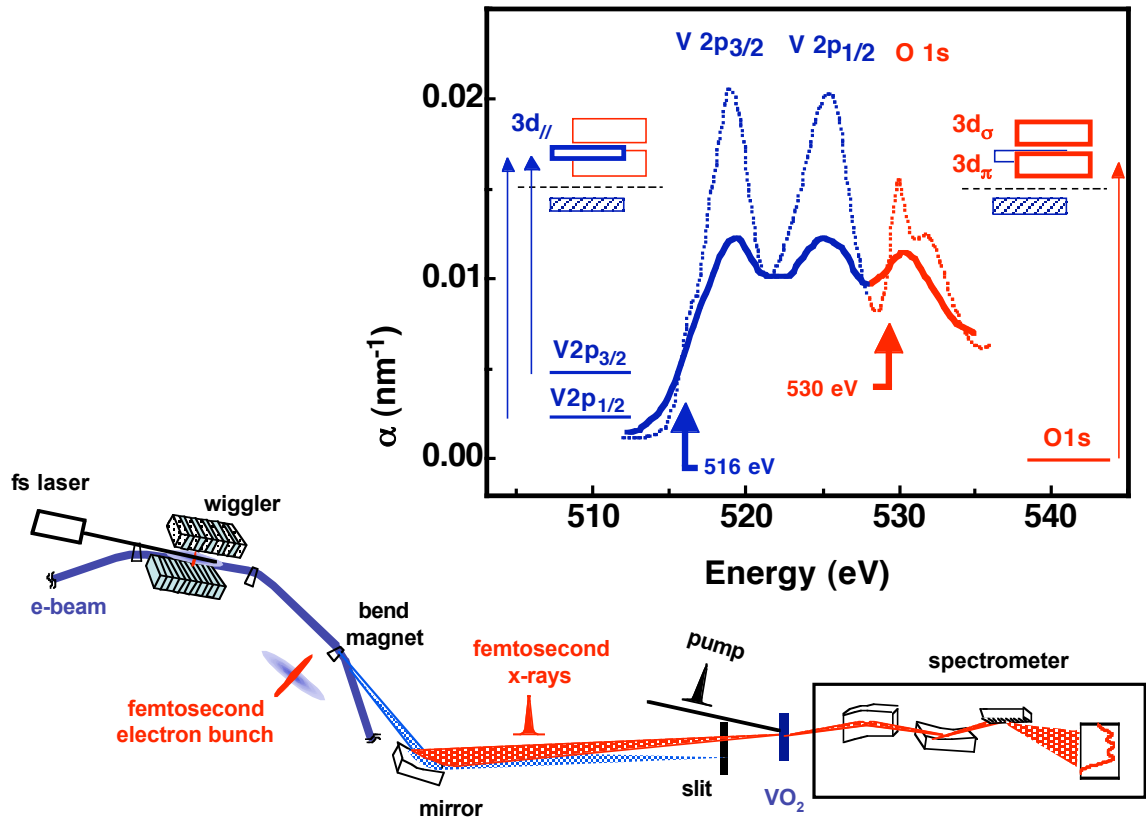

**Figure 2**

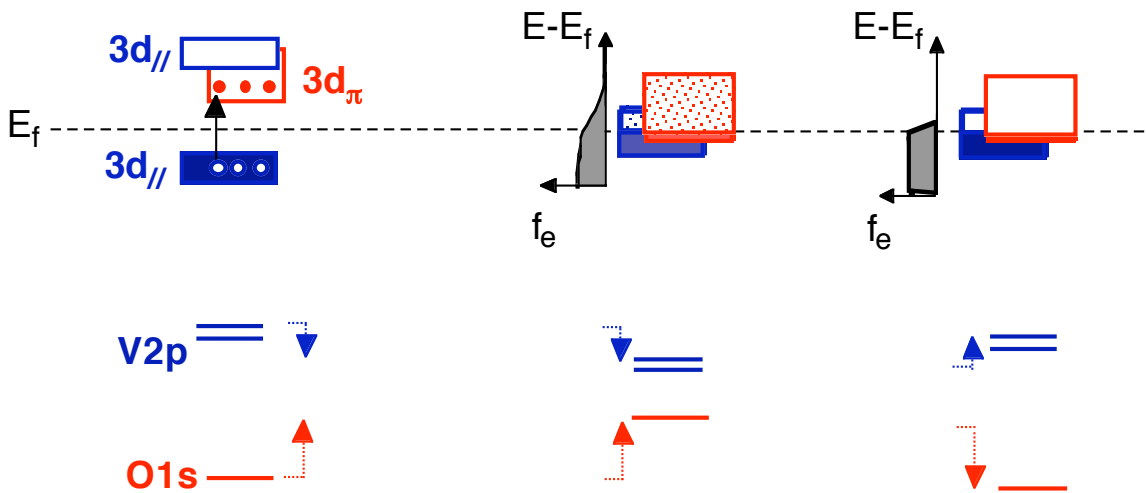
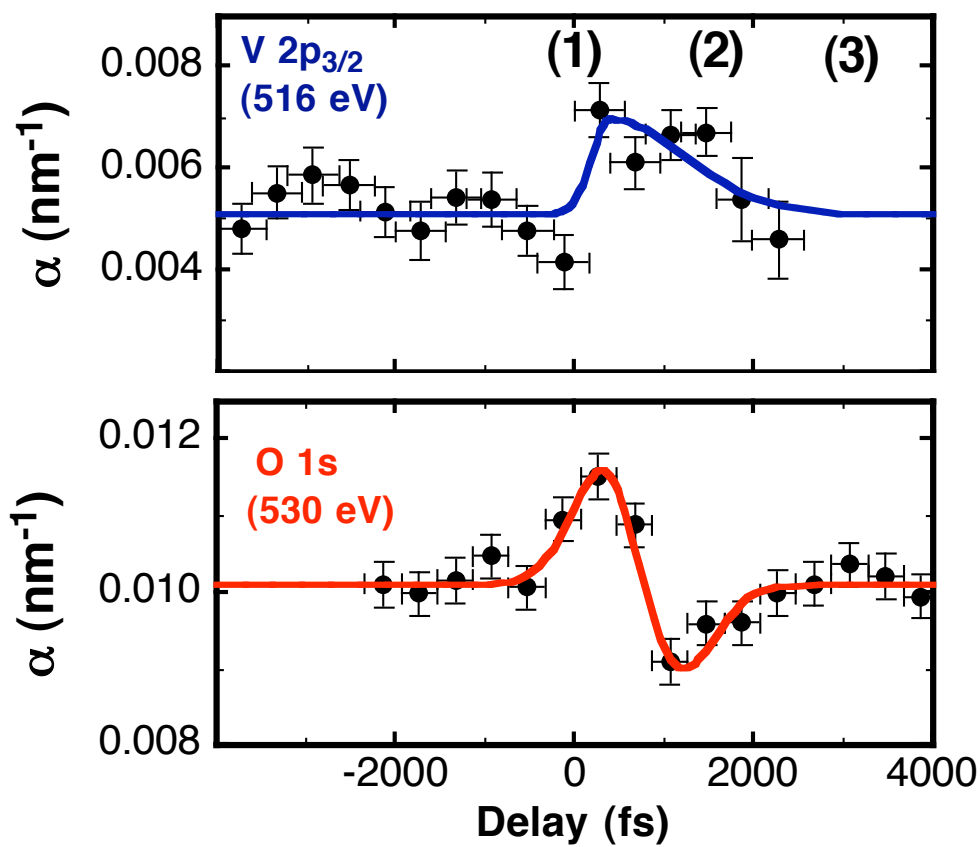

**Figure 3**